\def\vev#1{\langle #1 \rangle}

\def\ker{\mbox{Ker}\,}

\documentstyle[12pt]{article}
\begin{document}

\title{Fuzzy Space-Time}

\author{J. Madore \\
        Laboratoire de Physique Th\'eorique et Hautes 
        Energies\thanks{Laboratoire associ\'e au CNRS, \mbox{URA D0063}} \\ 
        Universit\'e de Paris-Sud, \
        B\^at. 211, F-91405 Orsay \\
       }

\date{}

\maketitle

\abstract{A review is made of recent efforts to define linear
connections and their corresponding curvature within the context of
noncommutative geometry. As an application it is suggested that it is
possible to identify the gravitational field as a phenomenological
manifestation of space-time commutation relations and to thereby
clarify its role as an ultraviolet regularizer.}

\vfill
\noindent
LPTHE Orsay 96/64
\medskip
\eject

\parskip 4pt plus2pt minus2pt

\section{Motivation}

Simply stated, noncommutative geometry is geometry in which the
`coordinates' do not commute. Since they cannot therefore be
simultaneously diagonalized, points are ill-defined. This is a fact
which is familiar from quantum mechanics; the canonical commutation
relations lead to the Heisenberg uncertainty relations.  The set of
functions on an ordinary space form a commutative algebra. Under certain
conditions it can be considered as an algebra generated by the
coordinates. In noncommutative geometry this algebra is replaced by a
noncommutative algebra which cannot be an algebra of functions; the
algebra remains but the space disappears. The geometry of ordinary
smooth spaces was written from the point of view of the algebra of
smooth functions by Koszul (1960) in his lectures at the Tata Institute.

There are three more-or-less independent arguments in favour of the use
of noncommutative geometry in high-energy physics and field theory. The
first is of a very practical nature.  When a physicist calculates a
Feynman diagram he is forced to place a cut-off $\Lambda$ on the
momentum variables in the integrands.  This means that he renounces any
interest in regions of space-time of dimension less than $\Lambda^{-1}$.
As $\Lambda$ becomes larger and larger the forbidden region becomes
smaller and smaller but it can never be made to vanish.  There is a
fundamental length scale, of the order of the Planck length, below which
the notion of a point is of no practical importance. The cut-off must be
bounded by the Planck mass.  The simplest and most elegant, if certainly
not the only, way of introducing such a scale in a Lorentz-invariant way
is through the introduction of non-commuting coordinates.

A closely related way of stating the same thing is to say that the
standard description of Minkowski space as a 4-dimensional continuum is
redundant. There are too many points. It was proposed already by
Heisenberg in the early days of quantum field theory to replace the
continuum by a lattice structure.  A lattice however breaks Poincar\'e
invariance and can hardly be considered as fundamental.  It was Snyder
(1947) who first had the idea of using non-commuting coordinates to
mimic a discrete structure in a covariant way. Since then several
attempts have been made to continue this initial effort. One typically
replaces the four Minkowski coordinates $x^\mu$ by four generators
$q^\mu$ of a noncommutative algebra ${\cal A}_{\mu_P}$ which satisfy
commutation relations of the form
$$
[q^\mu, q^\nu] = i \mu_P^{-2} q^{\mu\nu}.                         \eqno(1.1) 
$$ 
The problem lies then with the interpretation of the right-hand side.
The parameter $\mu_P$ is the Planck mass. One of our purposes is to
discuss the physical significance of $q^{\mu\nu}$ and its relation to
the gravitational field in the commutative limit. 

As a simple example of a space one can consider the ordinary round
2-sphere which has acting on it the rotational group $SO_3$. As a simple
example of a lattice structure one can consider two points on the
sphere.  One immediately notices of course that by choosing the two
points one has broken the rotational invariance and that however it can
be restored by admitting noncommuting `coordinates'. The set of
functions on the two points can be identified with the algebra of
diagonal $2 \times 2$ matrices, each of the two entries on the diagonal
corresponding to a possible value of a function at one of the two
points. Now an action of a group on a space is equivalent to an action
of the group on the algebra of functions on the space.  There can
obviously be no (non-trivial) action of the group $SO_3$ on the algebra
of diagonal $2 \times 2$ matrices but if one extends the algebra to the
noncommutative algebra of all $2 \times 2$ matrices one recovers the
invariance. The two points, so to speak, have been smeared out over the
surface of a sphere. They are replaced by two Bohr-like fuzzy regions.
Although what we have just done has nothing to do with Planck's constant
it is similar to the procedure of replacing a classical spin which can
take two values by a quantum spin of total spin 1/2. Only the latter is
invariant under the rotation group. By replacing the spin 1/2 by
arbitrary spin $s$ one can (Madore 1992) describe a `lattice structure' of
$2s+1$ points in an $SO_3$-invariant manner. On this structure the
notion of vectors and covectors can be introduced as well as the
noncommutative generalization of a metric and linear connection. It can
be readily shown that the unique torsion-free metric connection is
expressible in terms of the structure constants of the Lie algebra of
$SO_3$. We would like to examine to what extent this result can be
extended to noncommutative versions of Minkowski space, that is to what
extent the commutation relations which restrict the algebra determine
the possible gravitational fields which can be put on it.  By
`gravitational field' we shall mean linearized `gravity' in a
noncommutative version of Minkowski space. 

It is an old idea, due to Pauli and developed by Deser (1957) and others
(Isham {\it et al.} (1971), that perturbative ultraviolet divergences
will one day be regularized by the gravitational field. The possibility
which we would like to explore here is that the mechanism by which this
works is through the introduction of noncommuting `coordinates' such as
the $q^\lambda$. A hand-waving argument can be given (Madore \& Mourad
1995) which allows one to think of the noncommutative structure of
space-time as being due to quantum fluctuations of the light-cone in
ordinary 4-dimensional space-time.  This relies on the existence of
quantum gravitational fluctuations. A purely classical argument based on
the formation of black-holes has been given by Doplicher {\it et al.}
(1994, 1995).  In both cases the classical gravitational field is to be
considered as regularizing the ultraviolet divergences through the
introduction of the noncommutative structure of space-time. This can be
strengthened as the conjecture that the classical gravitational field
and the noncommutative nature of space-time are two aspects of the same
thing; they are both measures of the spectral densities of the operators
$q^\lambda$. The metric connection of the `fuzzy-sphere' example
mentioned above has a constant curvature and the spectral density of
each coordinate is uniform. One could say, inexactly but suggestively,
that the classical gravitational field arises from the `first
quantization' of the coordinates.  We use here the word `quantum' in the
loose way which implies that something does not commute. It has nothing
{\it a priori} to do with the `quantum' of quantum mechanics.

It is to be stressed that we modify the structure of space-time but
maintain covariance under the action of the Poincar\'e group. A fuzzy
space-time looks then like a solid which has a homogeneous distribution
of dislocations but no disclinations. We can pursue this solid-state
analogy and think of the ordinary Minkowski coordinates as macroscopic
order parameters obtained by coarse-graining over scales less than the
fundamental scale.  They break down and must be replaced by elements of
the algebra when one considers phenomena on these scales. Another
definition of `quantum space-time' implies that it is covariant under
the (co-)action of a `quantum' Poincar\'e group.

A linguistic digression is in order.  When referring to the version of
space-time which we describe here we use the adjective `fuzzy' to
underline the fact that points are ill-defined (Madore 1992, 1995).
Since the algebraic structure is described by commutation relations the
qualifier `quantum' has also been used (Snyder 1947, Doplicher {\it et
al.} 1995, Madore \& Mourad 1996b). This latter expression is
unfortunate since the structure has no immediate relation to quantum
mechanics and also it leads to confusion with `spaces' on which `quantum
groups' act.  To add to the confusion the adjective `fuzzy' has been
also used (Frittelli {\it et al.} 1996) to describe a rather different
noncommutative structure and the word `lattice' has been used ('t Hooft
1996) to designate what we here qualify as `fuzzy'.

In Section~2 we describe in more detail the basic algebra which we shall
suppose to replace the continuum description of Minkowski space. The use
of noncommutative geometry leads of course to many problems. One must
introduce the noncommutative generalization of a vector and a covector.
This is done in Sections~3 and 4. The gravitational field is
described by a metric and a linear connection.  In Section~5 the
noncommutative generalizations of these objects are recalled. There is
at the moment no completely satisfactory definition of the curvature of
a noncommutative linear connection in all generality but in the
particular cases which shall interest us here the usual definition in
terms of the covariant derivative will suffice. We are unable at the
moment to propose a satisfactory definition of an action and indeed we
are not in a position to argue that there is even a valid action
principle. A discussion of this point has been made by Connes and
coworkers in a series of articles (Kalau \& Walze 1995, Ackermann \&
Tolksdorf 1996, Connes 1996, Chamseddine \& Connes 1996) but the
definition which these authors propose is valid only on the
noncommutative generalizations of compact spaces with
euclidean-signature metrics. More technical details of the present
calculations are to be found in the article by Madore \& Mourad (1996b).

\section{Fuzzy space-time}

We start with an algebra ${\cal A}_{\mu_P}$ with 4 generators
$q^\lambda$ which satisfy the commutation relations (1.1). Since we wish
the limit space-time to be real we must suppose that the algebra has a
$*$-operation which replaces the complex conjugate and that the
$q^\lambda$ are hermitian. If we wished to do serious analysis we would
have to add a topology to ${\cal A}_{\mu_P}$.  Our first problem
is the interpretation of the right-hand side of (1.1).  It is the
value of the commutator $[q^\lambda, q^{\mu\nu}]$ which restricts the
structure of the algebra.  One possibility, considered by Snyder (1947),
is to choose it so that the algebra closes to form a representation of
the Lie algebra of the de~Sitter group.  A second possibility,
considered by Dubois-Violette \& Madore (Madore 1988, 1995) is to choose
it so that the algebra closes to form a representation of the conformal
algebra.  We refer to Madore \& Mourad (1996a) for a review with
historical perspective. Recently Doplicher, Fredenhagen \& Roberts
(1994, 1995) have argued that $q^{\mu\nu}$ should in fact be considered
as a new geometric quantity and that it can be chosen to lie in the
center of ${\cal A}_{\mu_P}$. We shall follow this suggestion here and
set therefore as a first approximation
$$
[q^\lambda, q^{\mu\nu}] = 0.                                      \eqno(2.1) 
$$
We shall see however that in order to introduce non-trivial
gravitational fields we shall be lead naturally to a non-vanishing term
on the right-hand side.  We shall suppose that the limit ${\cal A}_0$ of
the algebra ${\cal A}_{\mu_P}$ when $\mu_P \rightarrow \infty$ exists
and we set
$$
x^\lambda = \lim_{\mu_P \rightarrow \infty} q^\lambda.             \eqno(2.2)
$$
We shall see that the $x^\lambda$ are coordinates of an extension of
Minkowski space in the sense of Kaluza and Klein.

Not every space-time can be the limit of a noncommutative geometry.  
One can in fact define a Poisson structure on ${\cal A}_0$ by setting
$$
\{f,g\} = - i\lim_{\mu_P \rightarrow \infty} \mu^2_P [f,g].         \eqno(2.3)
$$
In particular 
$$
\{x^\mu, x^\nu\} = \lim_{\mu_P \rightarrow \infty} q^{\mu\nu}.
$$
The $q^{\mu\nu}$ cannot in the commutative limit tend to a set of
functions on space-time since this would break Lorentz invariance. 
(Any tensor except the metric breaks Lorentz invariance.) We
conclude quite generally then that space-time must be of dimension
greater than four if $q^{\mu\nu}$ does not identically vanish. Since
there are 6 independent $q^{\mu\nu}$ in the commutative limit one would
expect a space of at least 10 dimensions unless extra conditions are
imposed.  We shall avoid here the question of the physical significance
of the extra dimensions. We refer to Madore \& Mourad (1996a) for a
discussion of Kaluza-Klein theory within the context of noncommutative
geometry and to Dubois-Violette {\it et al.} (1996a) for a general
discussion of the commutative limits of noncommutative algebras.

We now turn to the Planck mass as a universal cut-off.  Let
$T^{(0)}_{\mu\nu}$ be the bare energy-momentum tensor, including quantum
corrections, of some field theory on space-time, Choose some separation
of $T^{(0)}_{\mu\nu}$ into a divergent part $T^{(\mu_P)}_{\mu\nu}$ and a
regular part $T^{(\mbox{\tiny Reg})}_{\mu\nu}$ which would remain finite
if one were to let $\mu_P \rightarrow \infty$.  Implicit in what follows
is the assumption that the decomposition can be made so that the
singular part is in some sense universal and independent of the
particular (physically reasonable) field theory one starts with. We
write then
$$
T^{(0)}_{\mu\nu} = T^{(\mu_P)}_{\mu\nu} + T^{(\mbox{\tiny Reg})}_{\mu\nu}.
                                                                    \eqno(2.4)
$$
Denote by $\vev{O}_0$ the vacuum-expectation value of an operator $O$. Then
in a quasi-classical approximation, considering the gravitational field
as classical, one can write the Einstein field equations as
$$
G_{\mu\nu} = - \mu_P^{-2} \big(\vev{T^{(\mu_P)}_{\mu\nu}}_0 +
                               \vev{T^{(\mbox{\tiny Reg})}_{\mu\nu}}_0\big).
$$
We shall be here interested in the divergent part of $T^{(0)}_{\mu\nu}$
and we shall neglect the regular term. The field equations become then
$$
G_{\mu\nu} = - \mu_P^{-2} \vev{T^{(\mu_P)}_{\mu\nu}}_0.           \eqno(2.5)
$$  
This equation is quite unsatisfactory. One would like to replace it by
an operator equation of the form
$$
G_{\mu\nu} = - \mu_P^{-2} T^{(\mu_P)}_{\mu\nu}                   \eqno(2.6)
$$  
such that
$$
G^{(\infty)}_{\mu\nu} = \lim_{\mu_P \rightarrow \infty} G_{\mu\nu}\eqno(2.7)
$$
is non-vanishing in order to produce a gravitational field which acts as
a regulator but such that
$$
\vev{G^{(\infty)}_{\mu\nu}}_0 = 0                                  \eqno(2.7)
$$
so that the regularizing gravitational field is not classically
observable.  In general however it is reasonable to assume that a
divergence gives rise to a gravitational field and with Equation~(2.5)
or (2.6) one can define the left arrow of the diagram
$$
\def\normalbaselines{\baselineskip=18pt}
\matrix{         &     \hbox{Cut-off}    &                  \cr
\hfill \swarrow  &                       & \nwarrow  \hfill \cr 
\hbox{Curvature} &    \longrightarrow    & \hfill \hbox{Algebra}
}                                                                  \eqno(2.9)
\def\normalbaselines{\baselineskip=12pt}
$$
The idea in introducing the noncommutative structure is to eliminate
points on small length scales. If this is achieved to within
$\mu^{-1}_P$ then one would expect that the algebraic structure
introduces a cut-off of order $\mu_P$. This is the right arrow of the
above diagram. It has been discussed, for example, by Doplicher {\it et
al.} The bottom arrow was discussed in the Introduction. It summerizes
the argument that a strong classical or quantum gravitational field
leads to a noncommutative version of space-time. 

To give more substance to this qualitative argument we examine the
following diagram:
$$
\def\normalbaselines{\baselineskip=18pt}
\matrix{         & \hbox{Calculus}  &                  \cr
\hfill \swarrow  &                  & \searrow  \hfill \cr 
\hbox{Curvature} & \longrightarrow  & \hfill \hbox{Algebra}
}                                                                  \eqno(2.10)
\def\normalbaselines{\baselineskip=12pt}
$$
We shall argue that it can in fact be used to define the bottom arrow,
the same as in (2.9). The right arrow is a mathematical triviality; it
gives a relation between a differential calculus (to be defined below)
over an algebra and the algebra itself. We shall argue that to a certain
extent a differential calculus determines uniquely a linear connection
in the commutative limit. The uniqueness will allow us to invert the
left arrow.  We can claim then that curvature gives rise not only to a
noncommutative algebra but also to an associated differential calculus
and as a corollary we have also defined the bottom arrow. To the extent
that this can be done we hope to use the construction of the curvature
in Diagram~(2.10) to replace the Equations~(2.5) or (2.6). That is,
instead of trying to use the field equations to deduce the curvature
directly from the divergent part $T^{(\mu_P)}_{\mu\nu}$ of
$T^{(0)}_{\mu\nu}$ we would like to envisage the possibility that it can
be indirectly inferred from the structure of the algebra which gives
rise to $T^{(\mu_P)}_{\mu\nu}$.  Since the differential calculus is not
unique we cannot claim that the curvature depends only on the right-hand
side of (1.1).  That is, although the non-vanishing Planck mass gives
rise to commutation relations, the left-hand side of (2.5) does not
depend only on $\mu_P$.  Were this the case then (2.5) would become an
eigenvalue equation yielding the mass spectrum in units of $\mu_P$.

\section{Differential calculi in general}

From the mathematical point of view a gravitational field is a rule for
displacing vectors and covectors and so we must have a definition of
these objects in the noncommutative case. We shall see that there are
certain problems associated with the definition of a vector but the
noncommutative generalization of a covector or differential form (Connes
1986, 1994) is quite satisfactory. A set of differential forms, with the
associated exterior derivative, is called a differential calculus.  The
main problem as far as physics is concerned is the fact that there is no
{\it a priori} unique way of defining a differential calculus over an
arbitrary algebra. In this section we shall give the general definitions
and illustrate them with simple examples. In the next section we shall
more specifically define differential calculi over the algebras defined
by relations of the form (1.1).

One of the characteristics of a noncommutative geometry is the lack of a
well-defined notion of localization. This is to be expected since there
are no points.  In the particular case of an algebra of functions on a
space one can speak of a function or element of the algebra as being
localized near a given point. Functions can be chosen whose support is
contained in an arbitrarily small region around the point. Vectors and
covectors can be studied in a given region of space without due
attention to whether or not their definition can be extended everywhere
on the space.  Noncommutative geometry on the other hand is essentially
global. When studying a smooth space from the point of view of
noncommutative geometry the algebra is the algebra of {\it all} smooth
functions. By vectors and covectors we mean therefore in this case {\it
globally defined} vector fields and covector fields.

There is a natural noncommutative generalization of a vector (field). To
see this we first consider the commutative case. Let $X^i$ be the
components of a vector and let $f$ be a smooth function. Then one can
form the derivative $X^i\partial_i f$ of $f$ in the direction $X^i$. 
It is again a smooth function. That is, one can consider
$X = X^i \partial_i$ as a linear map from the algebra of smooth functions
into itself which satisfies the Leibniz rule: $X(fg) = (Xf) g + f Xg$. A
map of this type of any (associative) algebra into itself is called a
derivation. So what we have shown is that to each vector field on a
smooth manifold one can associate a derivation of the algebra of smooth
functions on the manifold. It is easy to convince oneself that the
converse is true. We have therefore a natural generalization of a vector
field in the noncommutative case. It is a derivation of the algebra.
Notice that if $X$ is a derivation then so is $fX$ for any smooth
function $f$; the derivations form a left module over the algebra. 

As a simple noncommutative example it is instructive to consider again
the algebra $M_2$ of complex $2 \times 2$ matrices. Let $\lambda_i$ be
the Pauli matrices, chosen antihermitian and let $f$ be an arbitrary 
matrix. Then it is easy to see that the maps $e_i$ defined by
$$
e_i f = [\lambda_i, f]
$$ 
are derivations om $M_2$. Since $\lambda_i$ is antihermitian the
derivation is real; the matrix $e_i f$ is hermitian whenever $f$ is.  The
Leibniz rule is the Jacobi identity.  It can be shown that the most
general derivation of $M_2$ is of the form $X = X^i e_i$ where the $X^i$
are complex numbers. The set of derivations form a vector space of
dimension 3. This example serves to illustrate the fact that in the
noncommutative case a derivation cannot be multiplied from the left by
an arbitrary element of the algebra; the derivations do not form a left
module over the algebra. For this reason one usually tries to work as
much as possible in noncommutative geometry with the generalization of
covectors or differential forms.

Consider a smooth manifold and let $A_i$ be the components of a
(smooth) covariant vector (field). We shall write it as 
$A = A_i dx^i$ using a set of basis elements $dx^i$ and so
written we shall refer to it as a 1-form.  A 2-form is an antisymmetric
2-index covariant tensor $F_{ij}$ which we write as
$$
F = {1\over 2} F_{ij} dx^i dx^j
$$
using the product of the basis elements. This product is antisymmetric:
$$
dx^i dx^j = - dx^j dx^i                                            \eqno(3.1)
$$
but otherwise has no relations.  Higher-order forms can be defined as
arbitrary linear combination of products of 1-forms. A $p$-form can be
thus written as
$$
\alpha = {1\over p!} \alpha_{i_1 \cdots i_p} dx^{i_1} \cdots dx^{i_p}.
$$
The coefficients $\alpha_{i_1 \cdots i_p}$ are smooth functions and
completely antisymmetric in the $p$ indices. 

Let ${\cal A}$ be the algebra of smooth complex-valued functions on a
smooth manifold. We define $\Omega^0({\cal A}) = {\cal A}$ and for each
$p$ we write the vector space of $p$-forms as $\Omega^p({\cal A})$. Each
$\Omega^p({\cal A})$ depends obviously on the algebra ${\cal A}$ and,
what is also obvious and very important, it can be multiplied both from
the left and the right by the elements of ${\cal A}$.  It is easy to see
that $\Omega^p({\cal A}) = 0$ for all $p$ greater than the dimension of
the manifold.  We define $\Omega^*({\cal A})$ to be the set of all
$\Omega^p({\cal A})$.  We have seen that $\Omega^*({\cal A})$ has a
product. It is a graded commutative algebra.  It can be written as a sum
$$
\Omega^*({\cal A}) = \Omega^+({\cal A}) \oplus \Omega^-({\cal A})\eqno(3.2)
$$
of even forms and odd forms.  The algebra ${\cal A}$ is a subalgebra of
$\Omega^+({\cal A})$.

Let $f$ be a function, an element of the algebra 
${\cal A} = \Omega^0({\cal A})$.  We define a map $d$ from
$\Omega^p({\cal A})$ into $\Omega^{p+1}({\cal A})$ by the rules
$$
df = \partial_i f dx^i, \qquad d^2 = 0.
$$
It takes odd (even) forms into even (odd) ones. From the rules we
find that 
$$
d A = d(A_i dx^i) 
= {1\over 2}(\partial_i A_j - \partial_j A_i) dx^i dx^j = F
$$
if we set
$$
F_{ij} = \partial_i A_j - \partial_j A_i.
$$
From the second rule we have 
$$
d F = 0.
$$
It is easy to see that if $\alpha$ is a $p$-form and $\beta$ is a 
$q$-form then
$$
\alpha \beta = (-1)^{pq} \beta \alpha, \qquad
d(\alpha \beta) = (d \alpha) \beta + (-1)^p \alpha d\beta.
$$

The couple $(\Omega^*({\cal A}), d)$ is called a differential algebra or
a differential calculus over ${\cal A}$.  The algebra ${\cal A}$ need
not be commutative and $\Omega^*({\cal A})$ need not be graded
commutative. Over each algebra ${\cal A}$, be it commutative or not,
there can exist a multitude of differential calculi.  As a simple
example we define what is known as the universal differential calculus
$(\Omega_u^*({\cal A}), d_u)$ over a commutative algebra ${\cal A}$ of
functions. We set, as always, $\Omega_u^0({\cal A}) = {\cal A}$ and for
each $p \geq 1$ we define $\Omega_u^p({\cal A})$ to be the set of
$p$-point functions which vanish when any two points coincide. It is
obvious that $\Omega_u^p({\cal A}) \neq 0$ for all $p$.  There is a map
$d$ from $\Omega_u^p({\cal A})$ into $\Omega_u^{p+1}({\cal A})$ given in
the lowest order by
$$
(d_uf)(x, y) = f(y) - f(x).                                        \eqno(3.3)
$$
In higher orders it is given by a similar sort of alternating sum
defined so that $d_u^2 = 0$. The algebra $\Omega_u^*({\cal A})$ is not
graded commutative.  It is however defined for arbitrary functions, not
necessarily smooth, and it has a straightforward generalization for
arbitrary algebras, not necessarily commutative. To explain the qualifier
`universal' let $(\Omega^*({\cal A}), d)$ be any other differential
calculus over ${\cal A}$, for example the usual de~Rham differential
calculus constructed above. Then there is a unique $d_u$-homomorphism
$\phi$
$$
\Omega^*_u({\cal A}) \buildrel \phi \over \longrightarrow  \Omega^*({\cal A})  
$$
of $\Omega^*_u({\cal A})$ into $\Omega^*({\cal A})$. It is given by
$$
\phi (f) =  f,  \qquad \phi (d_u f) =  d f.
$$
If we choose a coordinate system and expand the function $f(y)$ about 
the point $x$,
$$
f(y) = f(x) + (y^i-x^i) \partial_i f + \cdots,
$$
we see that the map $\phi$ is given by
$$
\phi (y^i - x^i) = dx^i
$$
and that it annihilates any 1-form $f(x,y) \in \Omega^1_u({\cal A})$
which is second order in $x-y$. One such form is $fd_ug - d_ugf$,
given by
$$
(fd_ug -d_ugf)(x,y) = - \big(f(y) - f(x)\big)\big(g(y) - g(x)\big).
$$
It does not vanish in $\Omega^1_u({\cal A})$ but its image in 
$\Omega^1({\cal A})$ under $\phi$ is equal to zero.  For further
definitions we refer, for example, to the book by Connes (1994).  See
also Madore (1995).

What distinguishes the de~Rham differential calculus is the fact that it
is based on derivations. The relation between $d$ and $\partial_i$ 
is given by
$$
df (\partial_i) = \partial_i f.                                    \eqno(3.4)
$$
Notice that in particular this formula defines $dx^i$.  The
derivations form a vector space (the tangent space) at each point, and
the above equation defines $df$ as an element of the dual vector space
(the cotangent space) at the same point.  In the examples we shall
consider there are no points but the vector spaces of derivations are
still ordinary finite-dimensional vector spaces.  Over an arbitrary
algebra which has derivations one can always define in exactly the same
manner differential calculi based on the derivations.  These algebras
have thus at least two, quite different, differential calculi, the
universal one and the one based on the set of all derivations.

To illustrate the notion of forms in the noncommutative case we return
again to the algebra $M_2$ and show how to define $d\lambda_i$ using the
derivations (Dubois-Violette 1988, Dubois-Violette {\it et al.} 1989). 
Let $f$ be an element of $M_2$ and introduce the $SU_2$ structure
constants $C^i{}_{jk}$. We write them in this abstract form so that all
the formulae we write have immediate generalizations to the algebra
$M_n$ of $n \times n$ matrices provided we replace $SU_2$ by $SU_n$.
The straightforward extension of (3.4) can be written
$$
d\lambda_j (e_i) = [\lambda_i, \lambda_j] = C^k{}_{ij} \lambda_k.  \eqno(3.5)
$$
The set of $d\lambda_i$ constitutes a system of generators of
$\Omega^1(M_2)$ as a left or right module but it is not a convenient
one. For example $\lambda_i d\lambda_j \neq d\lambda_j \lambda_i$. There
is a better system of generators completely characterized by the
equations
$$
\theta^i(e_j) = \delta^i_j.                                        \eqno(3.6)
$$
We refer to the $\theta^i$ as a frame or Stehbein. They are
related to the $d\lambda_i$ by the equations
$$
d\lambda_i =  - C^j{}_{ik}\, \lambda_j \theta^k, \qquad
\theta^i = \lambda_j \lambda^i d\lambda^j.                         \eqno(3.7)
$$
They are the matrix analogue of the dual basis of the 1-forms. We have
raised and lowered indices here using the Killing metric on the Lie
algebra of $SU_2$. Because of the relation (3.6) we can write the
differential of any element $f$ of $M_2$ as
$$
df = e_i f \theta^i.                                               \eqno(3.8)
$$
Also because of the relation (3.6) we have
$$
\theta^i \theta^j = - \theta^j \theta^i, \qquad
\lambda_i \theta^j = \theta^j \lambda_i.                           \eqno(3.9)
$$
The product is the product in $\Omega^*(M_2)$; it is not in general
antisymmetric, Since the $\theta^i$ commute with the elements of $M_2$
we can identify $\Omega^1(M_2)$ with the tensor product of $M_2$ and the
dual $\bigwedge^1$ to the Lie algebra of $SU_2$:
$$
\Omega^1(M_2) = M_2 \otimes {\textstyle \bigwedge}^1.             \eqno(3.10)
$$
Technically speaking, $\Omega^1(M_2)$ is a free left (or right) 
$M_2$-module of rank equal to the dimension of the Lie algebra of $SU_2$. 
The geometry of $M_2$ is similar in this respect to the geometry of a
parallelizable manifold.

From the generators $\theta^i$ we can construct a 1-form
$$
\theta = - \lambda_i \theta^i                                   \eqno(3.11)
$$
in $\Omega^1(M_2)$ which satisfies the equation
$$
d \theta + \theta^2 = 0.                                        \eqno(3.12)
$$
It follows directly from the definitions that the exterior derivative
$df$ of an element of $M_2$ can be written in terms of a commutator with
$\theta$:
$$
df = - [\theta,f].                                               \eqno(3.13)   
$$
It follows that as a bimodule $\Omega^1(M_2)$ is generated by one element.

As a last example we consider a differential calculus over the algebra
of functions on a space of 2 points. We cannot use the construction
based on derivations since this algebra has none.  We write the algebra
$M_2$ as the direct sum
$$
M_2 = M^+_2 \oplus M^-_2
$$
of the diagonal matrices $M^+_2$ and the off-diagonal matrices $M^-_2$.
We saw in the introduction that we can identify the algebra of functions
on the space of 2 points with $M^+_2$. We set for each $p \geq 0$
$$
\Omega^{2p}(M^+_2) = M^+_2, \qquad \Omega^{2p+1}(M^+_2) = M^-_2.
$$
We choose an arbitrary (antihermitian)element $\theta$ in $M^-_2$ and we
define the differential $d$ by the formula (3.13) but with a graded 
bracket:
$$
[\alpha, \beta] = \alpha \beta - 
(-1)^{\vert\alpha\vert \vert\beta\vert} \beta \alpha. 
$$
Then we see immediately that if $\theta^2 = -1$ then $d^2 = 0$. One can
easily show that this differential calculus is in fact the universal
differential calculus over $M^-_2$. Although the space is rather trivial
the differential calculus has non-trivial entries in all dimensions.  We
can identify however all even and odd forms and write 
$\Omega^*(M^+_2) = M_2$. A formula similar to (3.13) can be written in
ordinary geometry using the Dirac operator instead of $\theta$ and a
differential calculus based on an appropriate generalization of (3.13)
can we defined for a large class of algebras (Connes 1994). 

To form tensors one must be able to define tensor products, for
example the tensor product 
$\Omega^1({\cal A}) \otimes_{\cal A} \Omega^1({\cal A})$ of 
$\Omega^1({\cal A})$ with itself.  We have here written in subscript the
algebra ${\cal A}$. This indicates the fact that we
identify $\xi f \otimes \eta$ with $\xi \otimes f \eta$ for every
element $f$ of the algebra and it means that one must be able to multiply
the elements of $\Omega^1({\cal A})$ on the left and on the right by the
elements of the algebra ${\cal A}$.  If ${\cal A}$ is commutative of
course these two operations are equivalent. When ${\cal A}$ is an
algebra of functions this left/right linearity is equivalent to the
property of locality. It means that the product of a function with a
1-form at a point is again a 1-form at the same point, a property which
distinguishes the ordinary product from other, non-local, products such
as the convolution. In the noncommutative case there are no points and
locality cannot be defined; it is replaced by the property of left
and/or right linearity with respect to the algebra.

The construction of the differential calculus over the algebra $M_2$
which we gave above relies on two properties of the derivations $e_i$.
First and most important these derivations are such that from the
identity $e_i f = 0$ follows that f must be proportional to the identity
matrix. Such elements are the noncommutative generalization of constant
functions. We have then `sufficient' derivations. The second property
which we would have used if we had entered into the details of the
construction of the higher-order forms is that the derivations form a
Lie algebra. It is in fact the Lie algebra of all derivations of $M_2$.
The first property is essential but this second property can be relaxed.
One can construct differential calculi over an algebra ${\cal A}$ based
on sets of derivations which form a proper Lie subalgebra of the Lie
algebra of all derivations or which indeed have no special Lie-algebra
property. To close this section we outline such a construction.

Let ${\cal A}$ be any algebra and let $(\Omega^*_u({\cal A}), d_u)$ be
the universal differential calculus over ${\cal A}$.  Every other
differential calculus can be considered as a quotient of the universal
one. For this construction we refer to the book by Connes (1994).  See
also Madore (1995).  Let $(\Omega^*({\cal A}), d)$ be another
differential calculus over ${\cal A}$. Then there exists a unique
$d_u$-homomorphism $\phi$ of $\Omega^*_u({\cal A})$ onto 
$\Omega^*({\cal A})$. It is given by
$$
\phi (d_u f) =  d f.                                               \eqno(3.14)
$$
The restriction $\phi_p$ of $\phi$ to each $\Omega^p_u$ is defined by
$$
\phi_p(f_0 d_u f_1 \cdots d_u f_p) = f_0 df_1 \cdots df_p.
$$

Consider a given algebra ${\cal A}$ and suppose that we know how to
construct an ${\cal A}$-module $\Omega^1({\cal A})$ and an application
$$
{\cal A} \buildrel d \over \longrightarrow \Omega^1({\cal A}).     \eqno(3.15)
$$
Then using (3.14) there is a method of constructing $\Omega^p({\cal A})$
for $p \geq 2$ as well as the extension of the differential.  Since we
know $\Omega_u^1({\cal A})$ and $\Omega^1({\cal A})$ we can suppose that
$\phi_1$ is given. We must construct $\Omega^2({\cal A})$. We shall
choose $\Omega^2({\cal A})$ to be the largest set of 2-forms consistent
with the constraints on $\Omega^1({\cal A})$. From general arguments we
know that it can be written in the from
$$
\Omega^2({\cal A}) = \Omega_u^2({\cal A}) / {\cal K}
$$
for some bimodule ${\cal K}$. Since we wish to have for every 1-form
$\xi_u$ in $\Omega_u^1({\cal A})$
$$
d \phi_1 (\xi_u) = \phi_2 (d_u \xi_u)
$$
we see that ${\cal K}$ must contain $d_u \ker \phi_1$. We can
choose ${\cal K}$ to be the bimodule generated by $d_u \ker \phi_1$.  
Let $\phi_2$ be the projection of $\Omega_u^2({\cal A})$ onto
$\Omega^2({\cal A})$.  The product of two elements $\xi$ and $\eta$ in
$\Omega^1({\cal A})$ is defined by choosing two inverse images $\xi_u$
and $\eta_u$ in $\Omega_u^1({\cal A})$ and projecting their product onto
$\Omega^2({\cal A})$:
$$
\xi \eta = \phi_2(\xi_u \otimes \eta_u).
$$
This procedure can be continued by iteration to arbitrary order in $p$.

To initiate the above construction we define the 1-forms using a set of
derivations. For each integer $n$ let $\lambda_i$ be a set of $n$
linearly independent antihermitian elements of ${\cal A}$ and proceed as
in (3.5) except that we can now write only
$$
d\lambda_j (e_i) = [\lambda_i, \lambda_j].
$$
To complete the construction we must postulate the existence
of a set of 1-forms $\theta^i$ which satisfy (3.6). In particular cases
this existence can be proven by explicit calculation. Notice that the
integer $n$ need not be equal to the number of generators of the algebra
but as a left or right module $\Omega^1({\cal A})$ is free of
rank $n$, that is, it has the $\theta^i$ as a basis over the
algebra. This is an essential property of parallelizable manifolds of
dimension $n$. Here the property can be made to hold for every integer
$n$, provided of course that the algebra ${\cal A}$ is noncommutative.
If a commutative limit is taken then the basis will be singular if $n$
is not equal to the dimension of the resulting manifold. We refer to
Dimakis \& Madore (1996) for an example of this. 

Because of the commutation relations of the algebra or, equivalently,
because of the kernel of $\phi_1$ in the quotient (3.15) the
$\theta^\alpha$ satisfy in general commutation relations. In one
important case which we shall consider they anticommute.
Because the 2-forms are generated by products of the $\theta^i$ 
one has
$$
d\theta^i = 
- {1\over 2} C^i{}_{jk} \theta^j \theta^k.                       \eqno(3.16) 
$$
The structure elements $C^i{}_{jk}$ can be {\it a priori}
arbitrary elements of the algebra.

Although we have made no explicit hypothesis concerning the $\lambda_i$
except that they be antihermitian and linearly independent, the existence
of the basis $\theta^i$ places severe restrictions on them. In fact one
can show that there must exist elements $P^{lm}{}_{jk}$, $F^i{}_{jk}$
and $K_{jk}$ in the center of the algebra such that
$$
2 \lambda_l \lambda_m P^{lm}{}_{jk} - \lambda_i F^i{}_{jk} - K_{jk} = 0.
                                                                  \eqno(3.17)
$$
The $P^{lm}{}_{jk}$ defines a projection of 
$\Omega^1({\cal A}) \otimes \Omega^1({\cal A})$ onto
$\Omega^2({\cal A})$. If it vanishes then so does $\Omega^2({\cal A})$.
From the associative rule for the algebra follow conditions on the
coefficients, Since the frame is determined by the derivations it is
to be expected that the structure elements are determined by the
coefficients in (3.17):
$$
C^i{}_{jk} = F^i{}_{jk} - 2 \lambda_l P^{(li)}{}_{jk}.             \eqno(3.18)
$$
We refer to Madore \& Mourad (1996b) for details.

\section{Differential calculi over fuzzy space-time}

The general formalism for the construction of differential calculi 
can be applied in particular to the algebra which we introduced in
Section~2. Suppose as a first approximation that (2.1) is satisfied and
that the matrix $q^{\mu\nu}$ has an inverse $q^{-1}_{\lambda\mu}$:
$$
q^{-1}_{\lambda\mu} q^{\mu\nu} = \delta^\nu_\lambda.
$$
We shall use this inverse to lower the indices of the generators $q^\mu$:
$$
\tilde q_\lambda = \mu_P^2 q^{-1}_{\lambda\mu} q^\mu.
$$
A natural choice of $n$ is $n=4$ and a natural choice of $\lambda_\mu$ 
is given by
$$
\lambda_\mu = - i \tilde q_\mu.                                     \eqno(4.1)
$$
The associated derivations defined in Section~2 satisfy then
$$
e_\mu q^\lambda = \delta^\lambda_\mu                                \eqno(4.2)
$$
and it follows that
$$
[e_\mu, e_\nu] = 0.                                                 \eqno(4.3)
$$
Notice that if $f$ is an element of ${\cal A}_{\mu_P}$ such that 
$e_\mu f = 0$ then we can only conclude that $f$ is an arbitrary 
function of the $q^{\mu\nu}$. We cannot conclude that it is proportional
to the identity. We accept this fact since we regard the non-trivial 
center as something which is to be eventually eliminated. With proper
conditions the center can be regarded as a smooth manifold and treated
using ordinary differential geometry.

From (4.2) it follows that
$$
\theta^\lambda = dq^\lambda, \qquad 
\theta = i \tilde q_\lambda dq^\lambda                              \eqno(4.4) 
$$
from which we deduce that
$$
P^{\mu\nu}{}_{\rho\sigma} = 
{1\over 2} (\delta^\mu_\rho \delta^\nu_\sigma - 
\delta^\nu_\rho \delta^\mu_\sigma), \qquad 
F^\lambda{}_{\mu\nu} = 0, \qquad
K_{\mu\nu} = i \mu_P^2 q^{-1}_{\mu\nu}.                             \eqno(4.5)
$$
From the commutation relations one finds that the $\theta^\lambda$
anticommute. One sees from their definition (3.16), or from (3.18),
that the structure elements vanish:
$$
C^\lambda{}_{\mu\nu} = 0.                                           \eqno(4.6)
$$
We shall see below in Section~5 that this implies that as in the
ordinary case the only torsion-free metric connection is the flat
one; the space-time is a noncommutative version of Minkowski space.
To find non-trivial gravitational fields we must change the differential
calculus. We would like to maintain the algorithm outlined in Section~2.
We must therefore either change the Ansatz (4.1) or the structure of the
algebra ${\cal A}_{\mu_P}$. 

We shall consider first-order perturbations of the $\lambda_\mu $
defined by (4.1).  Introduce four arbitrary `small' elements $f^\lambda$
of ${\cal A}$ and define
$$
\tilde f_\lambda = \mu_P^2 q^{-1}_{\lambda\mu} f^\mu.              \eqno(4.7)
$$
Then the elements 
$$
\lambda^\prime_\mu = - i (\tilde q_\mu +  \tilde f_\mu)            \eqno(4.8)
$$
are `near' to (4.1). In general, unless the condition (3.17) is
satisfied, $\Omega^2({\cal A}) = 0$ and the curvature will vanish.
Impose the condition (3.17) and let $P_{(1)}^{\rho\sigma}{}_{\mu\nu}$,
$F_{(1)}^\lambda{}_{\mu\nu}$ and $K_{(1)}{}_{\mu\nu}$ be the first-order
perturbations respectively of the coefficients.  By simple dimensional
arguments one can argue that $P_{(1)}^{\mu\nu}{}_{\rho\sigma}$ must
vanish. In fact it must tend to zero when the Planck mass tends to
infinity but on the other hand it is without dimension and therefore
cannot depend on the Planck mass.  Therefore we set
$$
P_{(1)}^{\mu\nu}{}_{\rho\sigma} = 0.                               \eqno(4.9)
$$
Using (4.5) we find that the linearization of (3.17) yields the equation
$$
[\tilde q_\mu, \tilde f_\nu] - [\tilde q_\nu, \tilde f_\mu] = 
i F_{(1)}^\lambda{}_{\mu\nu} \tilde q_\lambda - 
K_{(1)}{}_{\mu\nu}.                                               \eqno(4.10) 
$$
Let $k_\mu$ be an arbitrary `small' 4-vector with the dimension of mass.
Then a solution is given by
$$
\tilde f_\lambda = k_\mu q^\mu \tilde q_\lambda                   \eqno(4.11)
$$
and
$$
F_{(1)}^\lambda{}_{\mu\nu} = 
k^{\phantom{\lambda}}_{[\mu} \delta^\lambda_{\nu]} +
2 q^{-1}_{\mu\nu} q^{\lambda\sigma} k_\sigma, \qquad
K_{(1)}{}_{\mu\nu} = 0.                                            \eqno(4.12) 
$$
The corresponding frame is given by
$$
\theta^{\prime\lambda} = (1 - k_\rho q^\rho) dq^\lambda + 
q^{\lambda\rho} k_\rho q^{-1}_{\mu\sigma} q^\sigma dq^\mu.         \eqno(4.13)
$$
It will generate a new differential calculus 
$\Omega^{\prime *}({\cal A})$ which will be in general different 
from $\Omega^*({\cal A})$. Using (3.17) or (3.18) we find that 
$$
C_{(1)}^\lambda{}_{\mu\nu} = F_{(1)}^\lambda{}_{\mu\nu}.           \eqno(4.14)
$$
We shall see below that the modified calculus admits a gravitational
field. 

It is interesting also to maintain the Ansatz (4.1) but perturb the
condition (2.1) which restricted the algebra ${\cal A}_{\mu_P}$. We
introduce 6 `small' elements $q_{(1)}^{\mu\nu}$ of ${\cal A}_{\mu_P}$
and define
$$
q^{\prime\mu\nu} = q^{\mu\nu} + q_{(1)}^{\mu\nu}.                 \eqno(4.15)
$$
We have then
$$
q^{\prime -1}_{\mu\nu} = q^{-1}_{\mu\nu} + 
q^{-1}_{\mu\rho} q^{-1}_{\nu\sigma} q_{(1)}^{\rho\sigma}.         \eqno(4.16)
$$
Define
$$
\lambda^\prime_\mu = \lambda_\mu = - i \tilde q_\mu.              \eqno(4.17)
$$
We set
$$
[q^\lambda, q^{\prime\mu\nu}] = [q^\lambda, q_{(1)}^{\mu\nu}] = 
i \mu_P^{-1} q_{(1)}^{\lambda\mu\nu}.
$$
The simplest generalization of the condition (2.1) is to suppose that 
$q_{(1)}^{\lambda\mu\nu}$ lies in the center of ${\cal A}_{\mu_P}$. This
is the extended model of Doplicher {\it et al}. If we impose this
condition we can choose
$$
q_{(1)}^{\mu\nu} = - \mu_P^{-1} q_{(1)}^{\lambda\mu\nu} \tilde q_\lambda.
                                                                  \eqno(4.18)
$$
We have then
$$
e^\prime_\mu q^\nu = e_\mu q^\nu = \delta_\mu^\nu + 
\mu_P^{-1} q^{-1}_{\mu\rho} q_{(1)}^{\nu\rho\sigma} \tilde q_\sigma.
                                                                   \eqno(4.19)
$$
Using (4.5) we find that the linearization of (3.17) yields now the
solution
$$
F_{(1)}^\lambda{}_{\rho\sigma} = 
\mu_P q^{-1}_{\rho\mu} q^{-1}_{\sigma\nu}
\big(q_{(1)}^{\lambda\mu\nu} - q_{(1)}^{\mu\nu\lambda} + 
q_{(1)}^{\nu\mu\lambda}\big).                                       \eqno(4.20)
$$
The corresponding frame is given by
$$
\theta^{\prime\lambda} = dq^\lambda - \mu_P^{-1} q^{-1}_{\mu\rho} 
q_{(1)}^{\lambda\rho\sigma}\tilde q_\sigma dq^\mu.                  \eqno(4.21)
$$
From (4.14) we see that this leads again to a rather trivial but
nonvanishing gravitational field.

\section{Metrics and linear connections}

The definition of a connection as a covariant derivative was given an
algebraic form in the Tata lectures by Koszul (1960). We
shall use here the expressions `connection' and `covariant derivative'
synonymously. We shall consider it as a rule which to a covector 
$\xi = \xi_\lambda dx^\lambda$ associates its covariant derivative
$D\xi = D_\mu \xi_\nu dx^\mu \otimes dx^\nu$. For a detailed description
of the definition of a linear connection in noncommutative geometry we
refer to the article by Dubois-Violette {\it et al.} (1996b). We shall
here make use of the basis elements $\theta^\alpha$. That is,
since $\Omega^1({\cal A})$ is a free module a covariant
derivative can be defined by its action on these elements:
$$
D\theta^\alpha = 
- \omega^\alpha{}_{\beta\gamma} \theta^\beta \otimes \theta^\gamma.\eqno(5.1)
$$
The extension to arbitrary elements is given by the Leibniz rule. If
$\xi = \xi_\alpha \theta^\alpha$ is an arbitrary 1-form then
$$
D\xi = d\xi_\alpha \otimes \theta^\alpha + \xi_\alpha D\theta^\alpha.
                                                                   \eqno(5.2)
$$
The coefficients $\omega^\alpha{}_{\beta\gamma}$ are elements of the 
algebra.  Because of the identity 
$D(f \theta^\alpha) = D(\theta^\alpha f)$ they cannot however be 
arbitrary elements. To see this we rewrite $\xi$ as
$\xi = \theta^\alpha \xi_\alpha$ and we rewrite (5.2) as
$$
D\xi = \sigma(\theta^\alpha  \otimes d\xi_\alpha) + 
(D\theta^\alpha) \xi_\alpha.                                       \eqno(5.3)
$$
The purpose of the map $\sigma$ is to place the differential to the left
where is belongs while respecting the order of the terms. It is
discussed in detail in Dubois-Violette {\it et al.} (1996). In the
simple cases which we consider here it can be shown to be a simple
transposition. We can conclude therefore that the coefficients 
$\omega^\alpha{}_{\beta\gamma}$ must lie in the center of 
${\cal A}_{\mu_P}$. This is not always the case. 

One can define a metric by the condition
$$
g(\theta^\alpha \otimes \theta^\beta) = g^{\alpha\beta}           \eqno(5.4)
$$
where the coefficients $g^{\alpha\beta}$ are elements of the algebra.
To be well defined on all elements of the tensor product 
$\Omega^1({\cal A}) \otimes_{\cal A} \Omega^1({\cal A})$ the metric must
be bilinear and by the sequence of identities
$$
f g^{\alpha\beta} = g(f \theta^\alpha \otimes \theta^\beta) 
= g(\theta^\alpha \otimes \theta^\beta f) = g^{\alpha\beta} f     \eqno(5.5)
$$
one concludes that the coefficients must lie in the center of 
${\cal A}$.  This restriction plays an important role in restricting
the admissible connections. In the commutative limit the
$g^{\alpha\beta}$ cannot be functions of the coordinates. The Stehbein
not only determines the differential calculus it determines also
essentially the metric.

When $\sigma$ is a simple transposition the condition that $D$ be 
torsion-free becomes to first order the usual condition
$$
\omega^\lambda{}_{\mu\nu} - \omega^\lambda{}_{\nu\mu} = 
C^\lambda{}_{\mu\nu}.                                              \eqno(5.6)
$$
The condition that a connection be metric is a straightforward
generalization of the corresponding condition in the commutative case. 
Again when $\sigma$ is a simple transposition it becomes to
first order the condition
$$
\omega^\lambda{}_{\mu\nu} + 
\omega^{}_{\nu\mu}{}^\lambda = 0.                                 \eqno(5.7)
$$
The unique solution to these two conditions is given as usual by
$$
\omega^\lambda{}_{\mu\nu} = 
- {1\over 2} (C^\lambda{}_{\nu\mu} 
- C^{}_\mu{}^\lambda{}_\nu + C^{}_{\nu\mu}{}^\lambda).            \eqno(5.8)
$$
The basic calculus admits therefore no connections with non-vanishing 
curvature. However the two perturbations which we have considered in
Section~4 yield connections with non-vanishing albeit rather trivial
curvature in the commutative limit.  

There is no completely general satisfactory definition of the curvature
of a linear connection in the noncommutative case. However the map $D$
can be extended by
$$
\Omega^1({\cal A}) \otimes_{\cal A} \Omega^1({\cal A})
\buildrel D \over \longrightarrow 
\Omega^2({\cal A}) \otimes_{\cal A} \Omega^1({\cal A})           \eqno(5.9)
$$
and one might be tempted to define curvature by
$$
D^2 \theta^\alpha = - {1\over 2} R^\alpha{}_{\beta\gamma\delta} 
\theta^\gamma \theta^\delta \otimes \theta^\beta
$$
One can show that $D^2$ is left-linear,
$$
D^2 (f \theta^\alpha) = f D^2 \theta^\alpha,                      \eqno(5.10)
$$
but it is not in general right-linear:
$$
D^2 (\theta^\alpha f) \neq (D^2 \theta^\alpha) f.
$$
A detailed discussion has been given by Dubois-Violette {\it et al.}
(1996b). To lowest order we find the expression
$$
R^\mu{}_{\nu\rho\sigma} = 
\omega_{(1)}^\mu{}_{\rho\tau} \omega_{(1)}^\tau{}_{\sigma\nu} -
\omega_{(1)}^\mu{}_{\sigma\tau} \omega_{(1)}^\tau{}_{\rho\nu} -
\omega_{(1)}^\mu{}_{\tau\nu} C_{(1)}^\mu{}_{\rho\sigma}           \eqno(5.11)
$$
for the components of the curvature where the 
$\omega_{(1)}^\lambda{}_{\mu\nu}$ are determined in terms of the 
$C_{(1)}^\lambda{}_{\mu\nu}$ by (5.8). It is the `Einstein tensor' which
is obtained from this expression for the curvature which is to be placed
on the left-hand side of Equation (2.5). We shall not pursue this
further here since in order to make sense of the resulting equation an
average must be performed over the coordinates $q^{\mu\nu}$, 
$q^{\lambda\mu\nu}$ of the extra dimensions.

\section*{Acknowledgements} 

The author would like to thank J. Mourad for interesting conversations.

\parindent=0cm
\tolerance=1000
\parskip 5pt plus 1pt
\section*{References}

Ackermann T., Tolksdorf J. 1996, {\it A generalized Lichnerowicz
formula, the Wodzicki Residue and Gravity}, Jour. Geom. and Phys. (to
appear).

Chamseddine A., Connes A. 1996, {\it The spectral action principle} (in 
preparation). 

Connes A. 1986, {\it Non-Commutative Differential Geometry}, Publications
of the Inst. des Hautes Etudes Scientifique. {\bf 62} 257.

--- 1994, {\it Noncommutative Geometry}, Academic Press.

--- 1996, {\it Gravity coupled with matter and the foundation of non
commutative geometry}, IHES Preprint, hep-th/9603053.

Deser S. 1957, {\it General Relativity and the Divergence Problem in
Quantum Field Theory}, Rev. Mod. Phys. {\bf 29} 417.

Dimakis A., Madore J. 1996, {\it Differential Calculi and Linear 
Connections}, J. Math. Phys. (to appear).

Doplicher S., Fredenhagen K., Roberts, J.E. 1994, {\it Spacetime
quantization induced by classical gravity}, Phys. Lett. {\bf B331} 39.

--- 1995, {\it The Quantum Structure of Spacetime at the Planck Scale 
and Quantum Fields}, Commun. Math. Phys. {\bf 172} 187.

Dubois-Violette M. 1988, {\it D\'erivations et calcul diff\'erentiel 
non-commutatif}, C. R. Acad. Sci. Paris {\bf 307} S\'erie I (1988) 403.

Dubois-Violette M., Kerner R., Madore J. 1989, {\it Gauge bosons in a
noncommutative geometry}, Phys. Lett. {\bf B217} 485; {\it Classical
bosons in a noncommutative geometry}, Class. Quant. Grav. {\bf 6} 1709.

--- 1996a, {\it Shadow of Noncommutativity}, Preprint LPTHE Orsay 96/06.

Dubois-Violette M., Madore J., Masson T., Mourad J. 1996b, 
{\it On Curvature in Noncommutative Geometry}, 
J. Math. Phys. {\bf} (to appear).

Frittelli S., Kozameh C.N., Newman E.T., Rovelli C., Tate R.S. 1996,
{\it Fuzzy spacetime from a null-surface version of GR}, Pittsbourg
Preprint.

Isham C.J., Salam A., Strathdee J. 1971, {\it Infinity Suppression in 
Gravity-Modified Quantum Electrodynamics}, Phys. Rev. {\bf D3} 1805.

Kalau W., Walze M. 1995, {\it Gravity, Non-Commutative Geometry and 
the Wodzicki Residue}, Jour. Geom. and Phys. {\bf 16} 327.

Koszul J.L. 1960, {\it Lectures on Fibre Bundles and Differential Geometry},
Tata Institute of Fundamental Research, Bombay.

Madore J. 1988, {\it Non-Commutative Geometry and the Spinning
Particle}, XI Warsaw Symposium on Elementary Particle Physics,
Kazimierz, Poland; {\it Kaluza-Klein Aspects of Noncommutative
Geometry}, Proceedings of the XVII International Conference on
Differential Geometric Methods in Theoretical Physics, Chester.

--- 1992, {\it Fuzzy Physics}, Annals of Physics {\bf 219} 187.

--- 1995, {\it An Introduction to Noncommutative Differential Geometry
and its Physical Applications}, Cambridge University Press.

Madore J., Mourad. J. 1995, {\it On the origin of Kaluza-Klein
structure}, Phys. Lett. {\bf B359} 43.

--- 1996a, {\it Noncommutative Kaluza-Klein Theory}, Lecture given at the
$5^{\mbox{\tiny th}}$ Hellenic School and Workshops on Elementary
Particle Physics, hep-th/9601169.

--- 1996b, {\it Quantum Space-Time and Classical Gravity}, Preprint,
LPTHE Orsay, 95/56, gr-qc/9607060.

Mourad. J. 1995, {\it Linear Connections in Non-Commutative Geometry},
Class. Quant. Grav. {\bf 12} 965.

Snyder H.S. 1947, {\it Quantized Space-Time}, Phys. Rev. {\bf 71} 38.

't Hooft G. 1996 {\it Quantization of point particles in
(2+1)-dimensional gravity and spacetime discreteness}, 
Class. Quant. Grav. {\bf 13} 1023.

\end{document}